\documentclass[journal=jpccck,manuscript=article,layout=twocolumn]{achemso}
\setkeys{acs}{usetitle = true}

\usepackage[version=3]{mhchem} 
\usepackage[T1]{fontenc}       

\usepackage{amsmath,amssymb,bbm}
\usepackage{color}
\usepackage{lastpage}

\usepackage[normalem]{ulem}

\author{Joel Nicholls}
\affiliation{Department of Physics, University of Warwick, Coventry CV4 7AL, UK}
\alsoaffiliation{Centre for Complexity Science, University of Warwick, Coventry CV4 7AL, UK}
\email{Joel.Nicholls@warwick.ac.uk}
\author{Gareth P. Alexander}
\affiliation{Department of Physics, University of Warwick, Coventry CV4 7AL, UK}
\alsoaffiliation{Centre for Complexity Science, University of Warwick, Coventry CV4 7AL, UK}
\author{David Quigley}
\affiliation{Department of Physics, University of Warwick, Coventry CV4 7AL, UK}
\alsoaffiliation{Centre for Scientific Computing, University of Warwick, Coventry CV4 7AL, UK}
\email{D.Quigley@warwick.ac.uk}

\title{Polyomino Models of Surface Supramolecular Assembly: Design Constraints and Structural Selectivity}

\begin{document}


%
%
%
%



\begin{abstract}
We examine emergent properties of 2D supramolecular networks, using enumeration of configurations formed by interacting dominoes on
square lattices as a simple model system. Possible ground states are identified using a convex hull construction in the interaction parameters
for nearest-neighbour bonds. We demonstrate how this construction can be used to design interaction parameters which lead to networks with specific
properties, including chirality and highly degenerate ground states. We then introduce kinetics as simple local rearrangements. By partitioning
the configuration space into smaller sets which satisfy different topological constraints, we can design configurations which are kinetically trapped. 
By considering heat capacity curves along directions through the convex hull, we also demonstrate design of interacting domino configurations to create tilings  
robust against temperature induced phase transitions.
We discuss extension of this design construction to more complex molecular shapes.

\end{abstract}

\section{Introduction}
2D supramolecular self-assembly is a powerful tool for creating novel structures from the bottom up~\cite{whitelam15,whitesides02}. A range of systems can exhibit self-assembly given the right conditions: monolayers have been created at near-full packing density~\cite{beton10}, directional halogen bonds allow pyrene derivatives to form molecular networks on gold~\cite{pham14}, and ordered assemblies of millimetre size polymers at the perfluorodecalin-water interface~\cite{bowden97} are also possible via self-assembly. The necessary conditions for self-assembly of a given molecular network will depend on several different factors, for example the combination of both dynamics and hydrogen bonding is critical for an explanation of the assembly of 1,4-substituted benzenediamine nanostructures on gold substrate~\cite{haxton13}. Cyclohexa-m-phenylene will form different molecular networks, depending on whether it is adsorbed onto copper, silver, or gold substrates~\cite{bieri10}. It has been shown that changing the solvent, temperature, or molecular unit of a supramolecular network can alter the effective intermolecular interaction strength, leading to supramolecular networks with different structure, yet with molecular units of the same shape~\cite{betonAnother11}. Furthermore, it has been recently demonstrated that large chiral domains can be formed from achiral molecules~\cite{hu16}. Since many self-assembled systems have been shown to acquire useful or unusual properties, there is significant value in the rational design of self-assembled molecular networks.

A variety of techniques have been implemented to create self-assembled structures, such as deposition of molecules in ultra-high vacuum conditions, use of scanning tunelling microscopy to probe the structure, and the creation of regular molecular patterns via annealing~\cite{beton11,pham14}. Another avenue of self-assembly research is in DNA origami, where 2D crystalline arrays of origami tiles have been built~\cite{liu10}, and complex nano-scale shapes have been created out of self-assembled DNA strands~\cite{wei12}. Design principles have an advantage over the trial-and-error approach, for example it can be predicted that 1,4,5,6-naphthalenetetracarboxylic diimide-melamine adsorbed molecules will not assemble a honeycomb array because of the periodicity of the underlying silver-silicon substrate~\cite{beton11}. This kind of reasoning can be used to inform the choice of new experiments.

Aside from space-filling tilings, several studies have explored rational design of shapes and structures from site-specific interactions, or by entropic self-ordering~\cite{barnes09}. This includes complexity analysis of minimal sets of building blocks~\cite{ahnert10}, the use of patchy particles to form quasicrystals~\cite{doye12}, and self-assembly of charged soft dumbells~\cite{sciortino13}. Self-assembly principles are also critical for the production of many 3D structures, such as viral capsids~\cite{grime16} and supramolecular coordination complexes~\cite{cook13}. However, the design of specific interactions to assemble more complicated structures, say of lower symmetry, or with particular wallpaper groups~\cite{schattschneider78}, remains a significant challenge.

When molecular tilings are fully-packed, lattice models can be used effectively, since the molecules become interlocked. In particular, the statistical mechanics of dimer packings on the square lattice has been extensively studied since the work of Fisher~\cite{fisher61} and Kasteleyn~\cite{kasteleyn61,kasteleyn63}, who gave an exact expression for the partition function of the purely entropic model (zero interaction energy) in terms of Pfaffians and this work was expanded by others~\cite{rokhsar88,moessner01}. Subsequent studies have been primarily entropic or with achiral nearest neighbour aligning interactions~\cite{kundu14,ramola15}.

Theoretical tools can be used to describe the relevant behaviour of self-asembled molecular systems. Molecular dynamics simulations are used for realistic time evolution of self-assembly~\cite{martsinovich10}, partition functions encode the thermodynamics of assembled systems~\cite{jankowski11}, and investigation of assembly pathways provides new methods to generate ordered structures~\cite{nguyen11}. Furthermore, mathematical principles offer critical understanding, such as height functions to provide efficient computational sampling~\cite{korn04}, topological characterisation of the dynamic connectivity of states, and phase transition analysis to describe the temperature-dependence of system properties~\cite{kundu13}. However, prediction of experimental results remains a challenge. We propose a simple analogue model system for molecular networks and use it to find likely ordered configurations, and conversely, to determine conditions which would favour desired structures. An important method is the characterisation of possible energy assignments via nearest-neighbour interactions, which Istrail et al.~\cite{istrail00} have introduced to identify rhombus motifs on a hexagonal lattice. We use this method to show how to design for the symmetry group of the molecular network, degeneracy of ground state, robustness to entropic transitions, as dependent on the form of nearest-neighbour interactions, and in the presence of simple kinetics.

In this paper, we focus on the domino system, putting emphasis on the design of ground states with desired properties. Many of the methods generalise in a straightforward way for other polyomino shapes. Enumeration of configurations with various symmetries gives design principles for the wallpaper group of an ordered polyomino configuration, including the chirality of the polyomino packing. Degeneracy of the ground state is likewise an important quantity, since in experimental applications, a unique ground state with known properties is often preferred. Furthermore, we describe how to design for robustness of the ground state, both in terms of local kinetic moves (for dominoes), and for entropic phase transitions (for polyominoes).

\section{Methods}

To represent high-density packings of molecules adsorbed to a substrate, we use fully-packed polyomino configurations on a 2D square lattice. Furthermore, we consider edge-specific intermolecular interactions (see, for example, Figure~\ref{hull4by4}(a)) and describe the energy landscape in terms of the geometry of the state space. This identifies the lattice configurations that can be designed as ground states or as excited states. The nature of each ground state can also be used to indicate its robustness to an entropically driven phase transition. Our principle method is to create a complete construction of configuration space, and to represent it in a natural way.

The polyominoes interact through site specific interactions with energies $\varepsilon_{AB}$ between faces of type $A$ and $B$, so that the total energy of a particular configuration is determined by the number of interactions $n_{AB}$ of each type. A system made up of one kind of polyomino with $Q$ different face types has $Q(Q+1)/2$ different pairs of face types and hence this many interaction parameters $\varepsilon_{AB}$. Therefore, the general Hamiltonian for the system is of the form
\begin{equation}
H = \sum_{A=1}^{Q} \sum_{B\geq A}^{Q} n_{AB} \varepsilon_{AB}
\end{equation}
However, for a fully-packed system, there are packing constraints that reduce the effective dimension of the problem. There is only one type of polyomino in the system, and each face of each polyomino is in contact with a face of another polyomino. This gives a set of $Q$ linear equations relating the number of ``bonds'' to number of faces, which is in turn related to the number of plaquettes ($NM$ on an $N\times M$ lattice) of the periodic lattice containing that configuration
\begin{equation}
NM = n_{AA} + \sum_{B=1}^{Q} n_{AB}
\end{equation}
We can use these equations for each face type $A$, giving $Q$ equations to eliminate $Q$ interaction counts. All of the same-face interactions $n_{AA}$ can be eliminated to leave only $Q(Q-1)/2$ independent interaction parameters, after suitable redefinition of the parameters $\varepsilon_{AB}$. Similarly, configurations can be classified by a $Q(Q-1)/2$ component vector of interaction counts.

For the specific case of dominoes, and allowing for chiral interactions, there are 3 face types and, therefore, 6 different possible pairs of faces, but only 3 of the interaction counts are linearly independent, giving a 3D parameter space. We can choose our 3 interaction counts to be $n_{ab}, n_{ac}$ and $n_{bc}$ as shown in Figure~\ref{hull4by4}(a), thereby giving the chiral-interaction domino Hamiltonian

\begin{equation} \label{eq:Hamiltonian}
H = n_{ab} \varepsilon_{ab} + n_{ac} \varepsilon_{ac} + n_{bc} \varepsilon_{bc} .
\end{equation}

Individual configurations (indexed by $i$), can be labelled by the values of these interaction counts $\vec{n}^{i}$ and associated energy $E^{i} = \vec{n}^{i}\cdot \vec{\varepsilon}$. Each configuration therefore exists as a point in an abstract space ($\vec{n}$-space), with directions being the counts for each independent interaction. In general, there will be many distinct domino configuations with the same value of $\vec{n}$, meaning that they have equal energy under any choice of $\vec{\varepsilon}$. Using the Dancing Links algorithm~\cite{knuth00}, we have exhaustively enumerated all possible fully packed configurations of these dominoes on an $N\times M$ periodic square lattice, up to $N=M=8$.

In contrast to previous studies of tetromino fluids~\cite{barnes09,woszczyk15}, we are interested in the crystalline fully-packed configurations. The system of fully-packed dominoes has an equivalent representation as a height model~\cite{henley97,kenyon01}. In the height representation, adjacent vertices of the underlying square lattice are given integer values such that moving anticlockwise around an even plaquette will decrease the height by 3 when crossing a domino, and increase the height by 1 otherwise. This gives a unique representation for each domino configuration as a set of vertex heights, apart from a constant shift to the heights at all vertices. For a periodic domino configuration, the change in height has a mean value along both vertical and horizontal directions, which we will refer to as the height change per plaquette (a 2D vector). Furthermore, the coarse-grained properties of this height model can be used to show that for one set of interaction parameters on the infinite lattice~\cite{alet05}, the system undergoes a Berezinskii-Kosterlitz-Thouless-type phase transition.

For each configuration, we employ a discrete space algorithm to identify the wallpaper group of a tiling. This is based on the work in Ref.\citenum{schattschneider78}, however our algorithm is optimised for configurations which have an underlying lattice and is therefore ideal for polyomino tilings. Polyomino tilings on a square lattice cannot have hexagonal symmetry and, therefore, can be classified into one of 12 of the full 17 wallpaper groups.

The translational symmetry must include that of the original lattice, but may be larger. Any additional symmetries correspond to a subgroup of $\mathbb{Z}_N\times\mathbb{Z}_M$ and contain at least one subgroup of prime order. If a prime $p$ divides only one of $N$ or $M$, then there will be one subgroup with that order. If $p$ divides both $N$ and $M$, there will be $p\!+\!1$ subgroups. Checking such subgroups for each prime which divides $NM$ is the necessary and sufficient condition to check that there are no extra translational symmetries. This greatly increases the efficiency of the algorithm over a naive implementation. We also check for rotations, glides, or mirror symmetries. If we find there are no extra symmetries, the algorithm is finished, the wallpaper group is p1. If there are symmetries in the configuration, we first find the shortest two linearly independent translation vectors of the periodic tiling. This also tells us the lattice type of the pattern (square, rectangular, rhombic or oblique), and the size of the primitive cell. For each of the lattice types, there are a number of compatible wallpaper groups. This is identified by searching for the appropriate discrete symmetries.

\section{Results}

\subsection{Design of Crystalline Ground States}

The energy landscape of domino packings can be understood by visualising the set of all possible configurations in the $\vec{n}$-space. In this space, planes normal to the vector $\vec{\varepsilon}$ are necessarily iso-energetic. Extremal values of the scalar projection of $\vec{n}^{i}$ onto $\vec{\varepsilon}$ correspond to configurations with maximal or minimal energies. Hence configurations that lie on the boundary of the convex hull of points in this space represent packings that are ground states for an appropriately chosen $\vec{\varepsilon}$. The domino configurations that can be made unique energetic ground states are those corresponding uniquely to a vector $\vec{n}$ that is a vertex of the convex hull, provided this vertex corresponds to a single lattice configuration.

\begin{figure}[h]
\includegraphics[width=\linewidth]{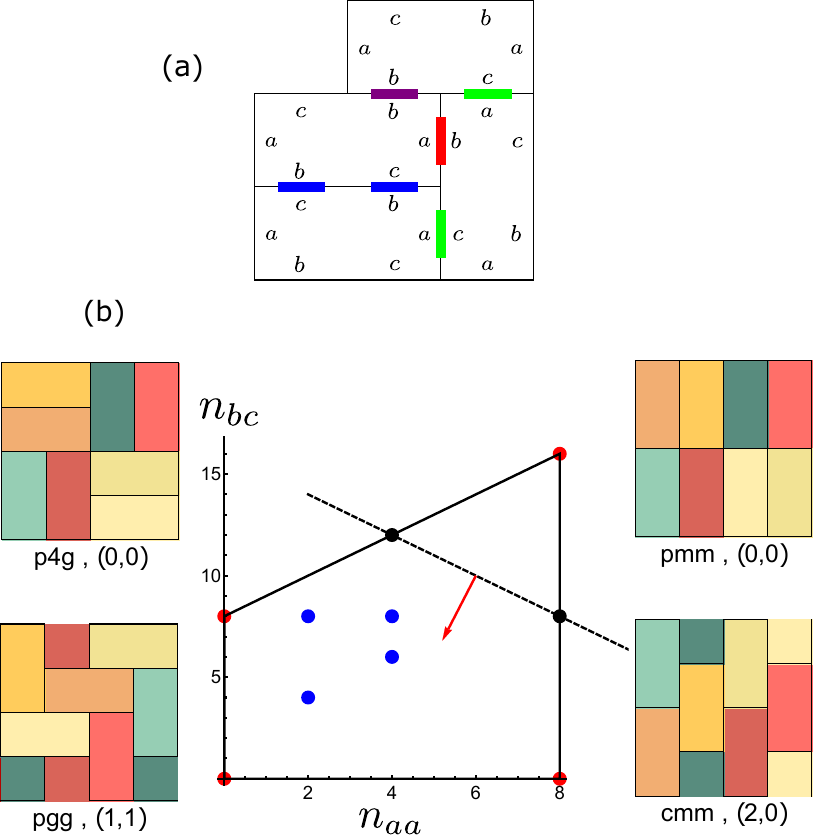}
\caption{\label{hull4by4}Configurations of dominoes on the $4 \times 4$ lattice. (a) Example domino collection, with faces $a$-$c$ labelled and coloured rectangles representing interactions between domino faces. (b) Points corresponding to configurations in $\vec{n}$-space, and the boundary of the convex hull is shown as dark lines. Vertex points are red, other boundary points are black, and interior points are blue. The red arrow gives an example vector of interaction parameters, such that the top-right configuration is the ground state. First excited states are given by the two points which the dashed line passes through. Also shown are the specific patterns for the vertex configurations, along with their wallpaper group and height change per plaquette.}
\end{figure}

An illustrative example is the $4\times 4$ lattice on which there are 13 possible configurations, some of which are displayed in Figure~\ref{hull4by4}(b). Of the full set of configurations, there is one with 4-fold rotational symmetry, 11 with 2-fold symmetry, and one without rotational symmetry. On the $4\times 4$ lattice the packings are further restricted by the condition $n_{ab}=n_{ac}=\frac{NM}{2}-n_{aa}$ so that the allowed configurations all lie in a 2D plane of the full configuration space. The hull is a quadrilateral so that there are four vertex ground states, depicted as red points in Figure~\ref{hull4by4}(b). The corresponding configurations happen to be unique to their position in $\vec{n}$-space, and have wallpaper groups pmm, p4g, pgg, and cmm (going anticlockwise from upper-right in Figure~\ref{hull4by4}(b)). These 4 configurations are therefore the possible non-degenerate ground states. For example, using a vector of interaction parameters $\vec{\varepsilon}=(-1,-1,-1)$, the p4g configuration can be designed as the ground state. This ground state has 4-fold rotation symmetry, primitive cell of area 8, and is non-degenerate for this combination of interaction parameters. However, if instead the interaction vector is $\vec{\varepsilon}=(-1,-1,-2)$, the ground state will be made up of 3 configurations, including the p4g configuraion, which is now degenerate under this choice of interaction parameters.

\begin{figure}[h]
\includegraphics[width=\linewidth]{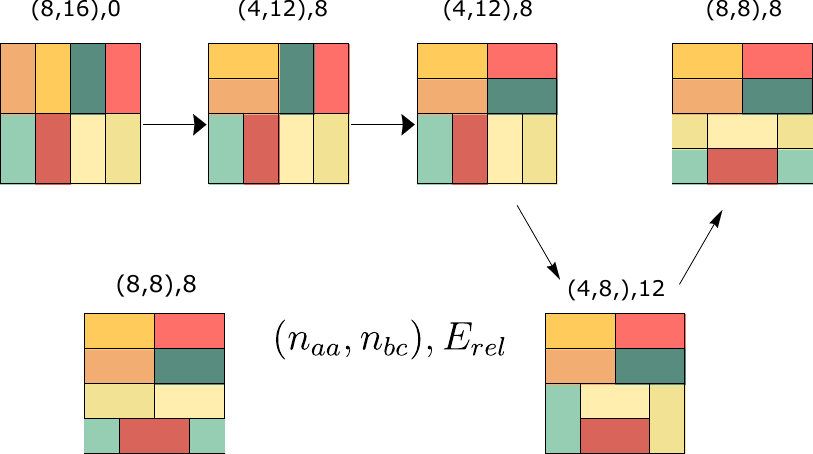}
\caption{\label{firstexcited}Sequence of local rearrangements of domino configurations on the $4 \times 4$ lattice, starting from the all-aligned configuration, and then moving to low energy excitations. Arrows indicate local rearrangements. $E_{rel}$ is the energy relative to the all-aligned configuration, under the choice $\varepsilon_{bc}=\varepsilon_{aa}=-1$ and all other components zero. This is the same as the example system given in Figure~\ref{hull4by4}, and here we visually give an example of an energy barrier for local rearrangement to a first excited state. Also, there is no set of local rearrangements that connects the lower-left configuration to the all-aligned configuration.}
\end{figure}

\subsection{Kinetically Restricted Design}

One can introduce kinetics into the model via local moves, which in this case takes the form of rotations of adjacent domino pairs (each of these moves is called a flip). As viewed from the height model representation, the height change per plaquette gives the criteria for domino configurations to be connected via flips, as described in Ref.\citenum{saldanha95}. For the four previously mentioned vertex states, the height change per plaquette is (respectively) $(0,0)$, $(0,0)$, $(1,1)$, and $(2,0)$, as shown in Figure~\ref{hull4by4}(b). To illustrate the concept of local moves, the domino pattern in the top left can be reached by the pattern in the top right by a sequence of flips. This sequence of local rearrangements is guaranteed to exist, since the two configurations have the same height change per plaquette of $(0,0)$. The orange arrow in Figure~\ref{hull4by4}(b) is an example vector for $\vec{\varepsilon}$, and in this case, the lowest value of $E = \vec{n} \cdot \vec{\varepsilon}$ corresponds to the configuration in the upper-right. Taking the vector of interaction parameters $\vec{\varepsilon}$ to have unit magnitude, the transition temperature to the higher entropy macrostate is $3.205$, in natural units. Furthermore, the two blue points on the boundary of the convex hull shown in Figure~\ref{hull4by4}(b) (close to the upper-right) constitute the first excited states of the Hamiltonian. These points are iso-energetic because they have equal scalar projection onto the example $\vec{\varepsilon}$ vector. Each of these two points corresponds to two domino configurations, but not all of these configurations can be reached from the ground state by local rearrangements~\cite{henley97}. Consequently, using only local moves, some configurations are completely inaccessible from the ground state, while others can only be reached by going through higher-energy configurations, as seen in Figure~\ref{firstexcited}.

For a domino configuration on a rectangular cell with periodic boundaries, not all configurations are connected by flips, which naturally gives a partition of configurations into sets of dynamically connected configurations. Given two dynamically connected configurations, there are generally several possible ways to make sequences of flips to transform from one configuration to the other. The transformation between them has an effective energy barrier as great as the lowest energy sequence of local rearrangements between them.

\begin{figure*}[t]
\includegraphics[width=\linewidth]{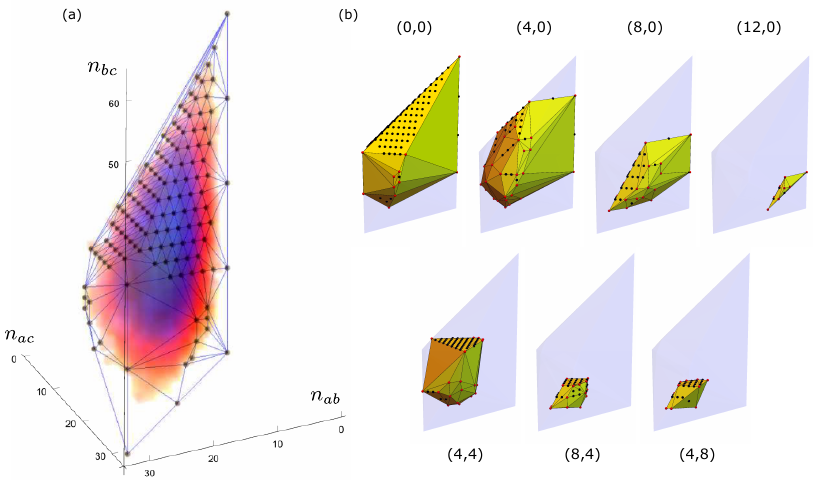}
\caption{\label{densityandconnections}(a) Points correspond to the configurations in $\vec{n}$-space which exist on the boundary of the convex hull of the set of configurations which can be made by dominoes on an $8\times 8$ lattice. The vertex points on the achiral plane are the same as for the $N=M=4$ case. Away from the achiral plane, the convex hull has 12 vertices corresponding to 24 configurations - 6 p1 and 18 p2. Configurations not on the achiral plane have cyclic (chiral) point groups. The line $n_{ab}=n_{ac},n_{bc}=0$ consists of the configurations with maximal $L^{1}$ norm of height change per plaquette, and the line $n_{ab}=n_{ac}=0$ contains the configurations which have only dominoes of one orientation. Also, the log of density of states is visualised as a heat map, with dark colours signifying high density. (b) The 7 non-trivial and inequivalent connected components of the $8\times 8$ lattice are shown in $\vec{n}$-space, with vertex points in red and other boundary points in black. A shadow is given to show the convex hull of the set of all domino configurations on the $8\times 8$ lattice. Alongside each of the 7 connected components is given the height change per plaquette.}
\end{figure*}

In addition to energetic design, it is possible to design ground state configurations that are kinetically isolated, meaning they cannot be converted into any other configuration via local moves, as they occupy a partition of size 1. For example, choosing $\vec{\varepsilon}=(-1,-1,1)$, we get the Herringbone pattern as unique ground state, which is shown in the lower-left of Figure~\ref{hull4by4}(b). For a system that allows only local rearrangements, this state will be topologically protected, and the system will not be able to explore the phase space of configurations. More generally, a given configuration will only be connected to some subset of the other configurations via local moves~\cite{saldanha95}. With the exception of the extremal case, each connected component is exactly the set of configurations with the same height change per plaquette. This separates configurations according to kinetic accessibility, as can be seen in Figure~\ref{densityandconnections}(b), where each connected component is depicted separately, for the $8\times 8$ lattice. If only local rearrangements are possible after adsorption, then the system is trapped in the connected component of the initial configuration, thereby giving the system a new set of possible effective ground states (shown in Figure~\ref{densityandconnections}(b)), which are not necessarily the true thermal ground state. This is a consequence of the fully-packed behaviour, which causes local moves to be non-ergodic in the full space of configurations.

\subsection{Chiral versus Achiral Configurations}

Configurations on the $4\times 4$ lattice are all achiral, in the sense that it is not possible to favour one configuration over its mirror image energetically. This a consequence of the additional condition $n_{ab}=n_{ac}=\frac{NM}{2}-n_{aa}$ on the number of interactions of each type. This is lifted for any larger square lattice and the set of possible configurations defines a 3D region in the $\vec{n}$-space. 

The convex hull for $N=M=8$, is shown in Figure~\ref{densityandconnections}(a), which has 1,224,518 configurations, divided into 1,551 different position vectors in $\vec{n}$-space. Packings that also arise on the smaller $4\times 4$ lattice lie in the plane $n_{ab}=n_{ac}$, a plane of mirror symmetry for all configurations. In comparison to the $4 \times 4$ lattice, the majority of configurations on the $8 \times 8$ lattice lie strictly within the boundary of the convex hull of points in $\vec{n}$-space. There are 12,237 boundary configurations, meaning that about 1\% of configurations of the $8\times 8$ lattice can be designed as ground state; of these configurations, only 4 achiral and 2 chiral pairs can be designed for as unique lowest energy states, (this is a strong limitation to the possible choices for the design of unique ground states).

Since we have an enumeration of states, we can design for desired properties by choosing the ground state which has the favoured properties. For example, on the $8 \times 8$ lattice, the vertex state $\vec{n} = (32,32,32)$  corresponds to exactly one configuration, which has 4-fold rotational symmetry. It is possible to design for 4-fold symmetry by choosing an appropriate combination of interaction energies to make this configuration the unique energetic ground state. One such choice would be $\vec{\varepsilon}=(-1,-1,-1)$, but generally, $\vec{\varepsilon}$ can be chosen as any solution to the set of linear inequalities which define the energy of the ground state to be lower than the energies of all other states. Conversely, one could design for a ground state with no rotational symmetry by finding a vertex state which corresponds to only configurations without rotational symmetry. For the $8 \times 8$ lattice, there are only two such vertex states: $\vec{n}=(34,20,24)$ and its chiral twin $\vec{n}=(20,34,24)$. The permitted periodicity of the configuration is important to this problem. For the $4 \times 4$ lattice, all ground states have some rotational symmetry, therefore it is not possible to design for a ground state without rotational symmetry in that case.

The degree of handedness~\cite{efrati14} of a domino packing can be given a qualitative value, using the distance from the achiral plane, defined as $|n_{ab}-n_{ac}|/\sqrt{2}$. To design an extremally chiral configuration, the vector of interaction energies should have the direction orthogonal to this plane, $\vec{\varepsilon}=(-1,1,0)$, so that the system will show a strong preference for one type of chiral interaction compared to the other.

This is not the only possible choice for the vector of interaction energies that gives rise to a chiral ground state. For the $8\times 8$ lattice, 13.8\% of the interaction parameter space corresponds to chiral ground states. However, to design for a non-degenerate chiral ground state, only 3.7\% of parameter space is available. This would require a greater control of the interaction parameters, which may be more challenging experimentally. 

To design for a very high degeneracy ground state, which is not simply the highest entropy macrostate, it is also necessary to have precise control of the interaction parameters. A striking feature of the boundary of the convex hull for $N=M=8$ are the faces with many distinct states, corresponding to ground states with high degeneracy for energy vectors $\vec{\varepsilon}$ perpendicular to either of these faces. There are two such faces, distinguished by their chirality. The equation for the normal of one of these faces is $NM=n_{ab}+n_{bc}$, meaning that the direction of the vector of interaction parameters has zero freedom, it must be chosen precisely. Also, it is possible to traverse the $\vec{n}$-space of this face by local rearrangements, meaning that this ground state is highly degenerate, even when considering kinetic constraints. However, it is not possible to reach all configurations which lie on this face, since there are some points that correspond to both locally accessible and inaccessible configurations.

\subsection{Robustness Against Temperature-Induced Phase Transitions}

\begin{figure}[h]
\includegraphics[width=\linewidth]{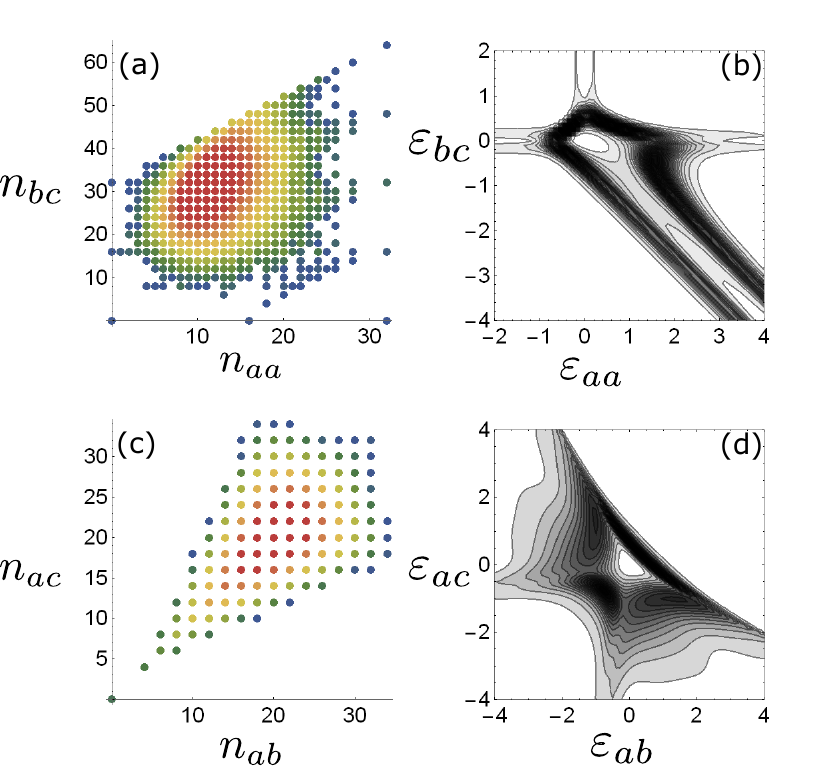}
\caption{\label{heatmapdos}The set of configurations considered here are the domino configurations which fit into an 8-by-8 lattice. For (a) and (b) we confine $\vec{\varepsilon}$ to the achiral plane, and for (c) and (d) we set $\varepsilon_{bc}=0$. Shown in (a) and (c) is the degeneracy of configurations, the colour scale goes from low (blue) to high (red), representing the logarithm of degeneracy. (b) and (d) give a heat capacity contour map, against energy interaction strengths, where the variation of energy with temperature is taken in the radial direction from the origin.}
\end{figure}

We now turn our attention to the selection of interactions that encode particular phase transitions. The set of parameters $\vec{\varepsilon}$ define a density of states in energy as the number of configurations associated to points $\vec{n}$, in planes orthogonal to the energy vector, which typically increases away from the extremal boundary configurations toward the centre of the convex hull (Figure~\ref{heatmapdos}(a) and (c)), leading to temperature induced phase transitions. A peak in heat capacity indicates the temperature at which this transition takes place. In fact, in the particular case of $\varepsilon_{bc}=-1$ and all other components zero, if the periodicity constraint is removed, it is already established that a Berezinskii-Kosterlitz-Thouless phase transition occurs~\cite{alet05}.

In the general case, one can map heat capacity as a function of the magnitude and direction of the $\vec{\varepsilon}$ vector. In Figure~\ref{heatmapdos}(b), heat capacity is shown as a function of the $\vec{\varepsilon}$ vector, where we have taken the chiral interaction energy $\varepsilon_{ab}-\varepsilon_{ac}$ to be zero. The four outer regions of Figure~\ref{heatmapdos}(b) correspond to the four vertex configurations, given by the vertices in Figure~\ref{heatmapdos}(a), which visualises the projection of the density of states onto the achiral plane. Importantly, the peak of the curve (as seen going radially outward) in Figure~\ref{heatmapdos}(b) gives an indication of the transition from ground state to the higher entropy macrostate.

We can design a ground state of high degeneracy by choosing interactions to be $\vec{\varepsilon}=(-1,-1,-2)$, thereby selecting the edge $n_{bc}=NM-(n_{ab}+n_{ac})/2$ as ground state, which has high degeneracy, as can be seen in Figure~\ref{heatmapdos}(a), since the points on the upper-left boundary have fairly high degeneracy. As shown in Figure~\ref{heatmapdos}(b), this choice corresponds to a relatively small heat capacity peak, which is between the parallel peaks in the lower-right quadrant. The transition away from the ground state occurs (for the $\vec{\varepsilon}$ given above, and natural units) at temperature 1.51104. This transition temperature is unusually high due to the high degeneracy of the ground state.

A further example is a transition that breaks chirality, as shown in Figure~\ref{heatmapdos}(d) for the subspace $\varepsilon_{bc}=0$. The upper-left and lower-right regions in Figure~\ref{heatmapdos}(d) are the left and right handed chiral ground states. The two other predominant regions of the phase diagram are the upper-right region, corresponding to all configurations of full orientational order, and the lower-left region, which corresponds to the collection of configurations with zero dominoes end-to-end. These regions indicate ground states which could be realised over a range of experimental parameter values.

The heat capacity peak in the chiral direction $\vec{\varepsilon}=(-1,1,0)$ is less sharp than a generic direction, due to there being several chiral states of similar energy, thereby giving a smoother transition between the (chiral) ground state and disordered state. The ground state established by $\vec{\varepsilon}=(-1,1,0)$ consists of 7 chiral configurations, and the transition away from this ground state occurs at temperature 0.787068 (in natural units).

\begin{figure}[h]
\includegraphics[width=0.8\linewidth]{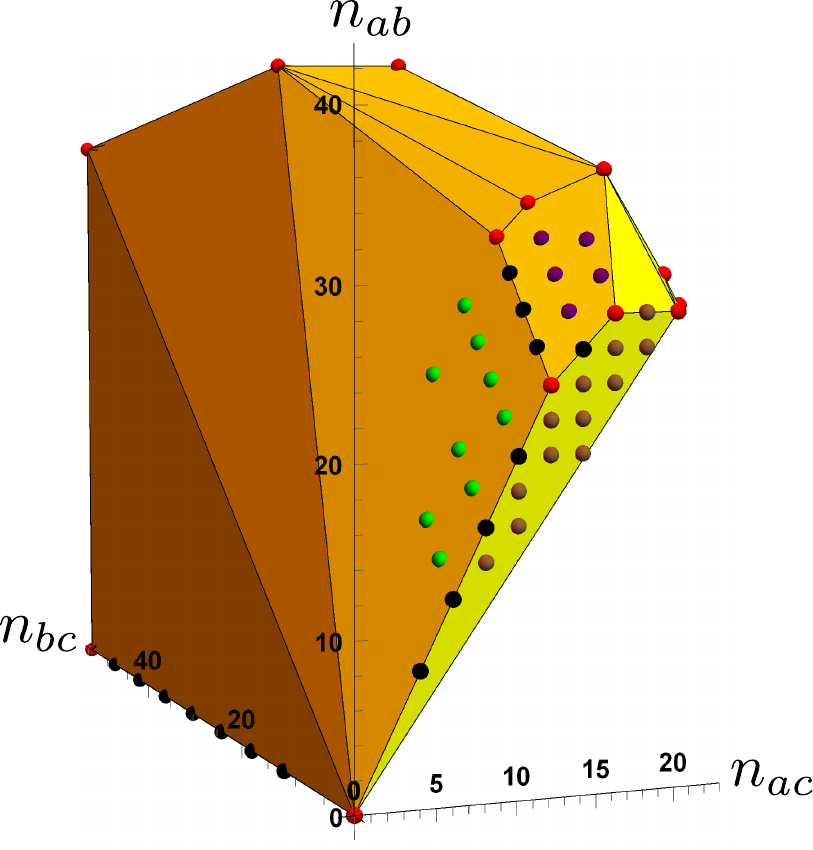}
\caption{\label{polyominofigure}The convex hull boundary of the set of linear trimers that can be made on a $9\times 9$ lattice, plotted in $\vec{n}$-space for non-chiral interactions. Red points show vertices and black points are boundary points of the convex hull. The colours purple, green and brown have been used to indicate points on 3 different faces of the convex hull.}
\end{figure}

Keeping the above examples in mind, some general rules become apparent for the design of ground states that are robust to temperature-induced phase transitions. When the vector of interaction parameters has greater magnitude, the interactions are overall stronger, which raises the transition temperature to the high-entropy macrostate. Also important is the direction of the vector of interaction parameters, since this will determine the smoothness of the transition from the ground state to the high-entropy macrostate, as well as influencing the temperature at which this transition occurs. Robustness against transition to the high-entropy macrostate is also strongly affected by the kinetic restrictions we mentioned previously. Due to these kinetic restrictions, the ground state configurations with greater height change per plaquette are able to access a smaller set of excited states via local rearrangements. In particular, for configurations with extremal height change per plaquette, there is no possibility for rearrangement via domino pair flips.

\subsection{Polyominoes}

Many of the methods and techniques used here will translate directly to other kinds of polyomino, for example longer rectangular polyominoes~\cite{kundu14} or T-tetrominoes~\cite{korn04}. For general polyomino shapes a convex hull construction can be used to identify possible ground states. As a brief illustration we show in Figure~\ref{polyominofigure} the convex hull for non-chiral linear trimers. Packing constraints have been used to reduce the effective dimension of the problem and properties of the configurations have been computed, such as the wallpaper groups, using direct generalisations of methods from the domino system. Similar to domino configurations, we find that for linear trimers, there are possible ground states with especially high degeneracy. These states correspond to highly populated faces of the convex hull in $\vec{n}$-space and, again, it is possible to find local move sets which traverse the faces.

Specifically, for linear trimers on the $9\times 9$ lattice, three highly degenerate faces of the convex hull meet at one vertex state. The three faces are shown by differently coloured individual states on each face, in Figure~\ref{polyominofigure}. All three faces are close to being parallel, such that only 1.28\% of interaction parameter space lies in the region between the three highly degenerate states. This means that each of these faces can be realised as lowest energy state under similar interaction parameters, resulting in sensitive dependence on the form of the nearest-neighbour interactions. Each of these strongly degenerate ground states have high transition temperatures to the high-entropy macrostate, analogous to the domino system. For example, the vector of interaction parameters $\vec{\varepsilon}=(0,0,1)$ corresponds to a highly degenerate ground state with a transition temperature at 1.50625, whereas the choice $\vec{\varepsilon}=(0,-1,0)$ corresponds to a low degeneracy ground state of transition temperature 0.73366. 

In the case of linear trimers, there is also a concept of a height function~\cite{jacobsen07}. However, the question of connectivity of configurations via local moves is not as straightforward as for dominoes.

\section{Discussion and Conclusions}

We have shown how properties of periodic 2D supramolecular networks can be selected in terms of nearest-neighbour interactions between individual molecules. We describe details of the relationships between the properties of the molecular network that forms and the parameters that brought it about. Furthermore, we demonstrate how these relationships can be used as design principles for informing the conditions needed to create a given desired property. A range of properties have been described in terms of design, such as: degeneracy and symmetry of ground state, chirality, kinetic trapping, and entropic robustness. These properties depend strongly on the type of nearest-neighbour interactions, motivating visualisation of the set of configurations in $\vec{n}$-space as a design tool. In particular, we have found the possibility for high-degneracy and kinetically trapped ground states. Many experiments have previously demonstrated ordered domains at very high packing fraction~\cite{hu16,bieri10,beton11,pham14,betonAnother11}, and our results indicate new directions in which to take these kinds of experiment.

There are many directions for future work. We have only demonstrated fully-packed configurations, for which there already exists a significant literature on the corresponding lower-density systems. The calculation of a convex hull to determine ground states becomes much more difficult in higher dimensions, which restricts the number of different interaction types we can meaningfully consider simultaneously. An approach based on sampling of periodic configurations, rather than enumeration, could bring greater insight. For an algorithm performing depth-first enumeration, it is possible to modify the algorithm to perform Markovian dynamics on the partially completed configuration space, such that completed configurations are sampled with equal probability. This would make it possible to investigate configurations of larger periodicity. 

The principles and methods described here will hopefully provide inspiration for the design of self-assembling monolayers. In this paper, we have kept primarily to domino configurations, but the methods presented are general for systems of any other type of polyomino. However, only some polyominoes have a currently known height function~\cite{pak00,jacobsen07,korn04}. For those that do have a height function, this does not always give a sufficient condition for configurations to be connected by local moves. Nonetheless, it may be possible to use those height functions to look for necessary conditions for connectivity by local moves. Another complication related to general polyominoes, is that the $\vec{n}$-space of other polyominoes can be higher than 3D, making the convex hull more difficult to visualise. In terms of design, we would like to stress that it is the convex hull of configurations in $\vec{n}$-space, which gives the relevant information on the possible ground states of the system, under very general nearest-neighbour interactions.

\begin{acknowledgement}

This work was partially supported by the UK EPSRC through Grant No. EP/I01358X/1 (JN and GPA).

\end{acknowledgement}

%
%
%


\providecommand{\latin}[1]{#1}
\makeatletter
\providecommand{\doi}
  {\begingroup\let\do\@makeother\dospecials
  \catcode`\{=1 \catcode`\}=2\doi@aux}
\providecommand{\doi@aux}[1]{\endgroup\texttt{#1}}
\makeatother
\providecommand*\mcitethebibliography{\thebibliography}
\csname @ifundefined\endcsname{endmcitethebibliography}
  {\let\endmcitethebibliography\endthebibliography}{}


\end{document}